\documentclass[lettersize,journal]{IEEEtran}
\usepackage{amsmath,amsfonts}
\usepackage{algorithmic}
\usepackage{algorithm}
\usepackage{array}
\usepackage[caption=false,font=normalsize,labelfont=sf,textfont=sf]{subfig}
\usepackage{textcomp}
\usepackage{stfloats}
\usepackage[normalem]{ulem}
\usepackage{url}
\usepackage{verbatim}
\usepackage{graphicx}
\usepackage{cite}
\usepackage{cite}
\usepackage{amsmath,amssymb,amsfonts}
\usepackage{xcolor}
\usepackage{multirow}
\usepackage{tabularx}
\usepackage{booktabs}
\usepackage{makecell}
\usepackage[colorlinks=true, allcolors=blue]{hyperref}

\newcount\Comments  
\usepackage{color}
\definecolor{darkgreen}{rgb}{0,0.5,0}
\definecolor{purple}{rgb}{1,0,1}
\newcommand{\kibitz}[2]{\ifnum\Comments=0\textcolor{#1}{#2}\fi}

\begin{document}

\title{Real-World Deployment and Assessment of a Multi-Agent Reinforcement Learning-Based Variable Speed Limit Control System}


\author{Yuhang Zhang$^{\dagger*}$, Zhiyao Zhang$^{\dagger}$, 
Junyi Ji$^{\dagger}$, Marcos Qui\~{n}ones-Grueiro$^{\dagger}$, William Barbour$^{\dagger}$, Derek Gloudemans$^{\dagger}$, Gergely Zachár$^{\dagger}$, Clay Weston$^{\ddagger}$, Gautam Biswas$^{\dagger}$, Daniel B. Work$^{\dagger}$
\thanks{$^{\dagger}$Institute for Software Integrated Systems, Vanderbilt University, Nashville TN, 37212, USA. $^{\ddagger}$Southwest Research Institute, San Antonio, TX, 78238, USA. $^*$yuhang.zhang.1@vanderbilt.edu}
}



\maketitle

\begin{figure*}
    \centering
    \includegraphics[width=2\columnwidth]{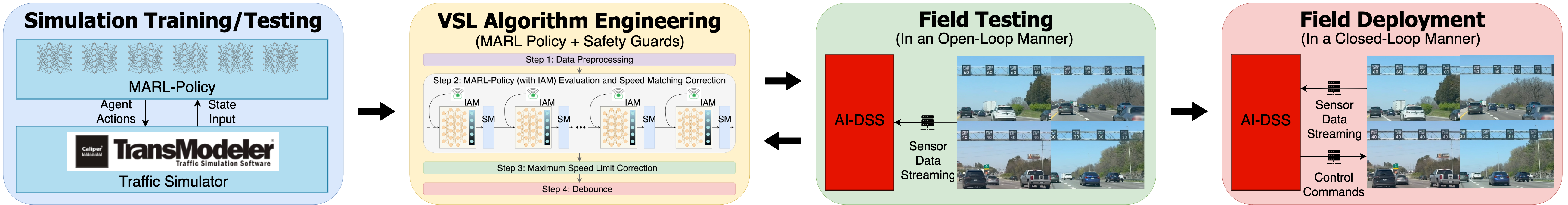}
    \caption{Deployment pipeline of our MARL-based VSL control system: \textit{\textbf{Step 1:}} We trained 8 agents in a traffic micro-simulation software TransModeler on a 7-mile stretch of I-24 and then tested it with 34 agents on a 17-mile stretch of westbound I-24 with various simulation parameters. \textit{\textbf{Step 2:}} We extracted the optimal policy learned from simulation and applied invalid action masking and safety guards to satisfy real-world constraints. \textit{\textbf{Step 3:}} We tested the behavior of the proposed MARL-based VSL control algorithm in an open-loop manner, with continuous streaming of I-24 sensor data feeding into Artificial-Intelligence Decision Support System (AI-DSS), the infrastructure software served for communication with \textit{Traffic Management Center} (TMC). Based on the testing results, we go back to Step 2 to refine our algorithm until it presents satisfying behaviors. \textit{\textbf{Step 4:}} We deployed the MARL-based VSL control algorithm in a closed-loop manner across 67 VSL controllers spanning a 17-mile segment of I-24 on March 8, 2024. The MARL-based VSL control system is continuously operating on I-24 today, affecting nearly 160,000 daily commuters.}
    \label{fig:pipeline}
\end{figure*}

\begin{abstract}
This article presents the first field deployment of a multi-agent reinforcement learning (MARL) based variable speed limit (VSL) control system on Interstate 24 (I-24) near Nashville, Tennessee. We design and demonstrate a full pipeline from training MARL agents in a traffic simulator to a field deployment on a 17-mile segment of I-24 encompassing 67 VSL controllers. The system was launched on March 8th, 2024, and has made approximately 35 million decisions on 28 million trips in six months of operation. We apply an invalid action masking mechanism and several safety guards to ensure real-world constraints.  The MARL-based implementation operates up to 98\% of the time, with the safety guards overriding the MARL decisions for the remaining time. We evaluate the performance of the MARL-based algorithm in comparison to a previously deployed non-RL VSL benchmark algorithm on I-24. Results show that the MARL-based VSL control system achieves a superior performance. The accuracy of correctly warning drivers about slowing traffic ahead is improved by 14\% and the response delay to non-recurrent congestion is reduced by 75\%. The preliminary data shows that the VSL control system has reduced the crash rate by 26\% and the secondary crash rate by 50\%. We open-sourced the deployed MARL-based VSL algorithm at https://github.com/Lab-Work/marl-vsl-controller. 
\end{abstract}


\section{Introduction}  
\label{sec:introduction}

As urbanization intensifies and vehicle ownership continues to rise, traffic congestion and the frequency of traffic incidents have emerged as significant challenges in the transportation field~\cite{FHWA_statistics}. Since expanding roadway capacity through additional lanes is both physically and economically challenging, much attention has been paid to traffic management strategies that better utilize existing infrastructure. Among these strategies, \textit{variable speed limit} (VSL) control works by regulating the mainline highway traffic through dynamic speed limit adjustments based on real-time traffic conditions, thereby mitigating congestion and reducing crashes~\cite{lu2014review, khondaker2015variable}. Concurrently, the recent availability of comprehensive traffic data and advanced \textit{artificial intelligence} (AI) techniques provide possibilities for more advanced, data-driven algorithms~\cite{10.1145/3543507.3583452}.


Emerging from advancements in AI, \textit{reinforcement learning} (RL) provides a framework for adaptive decision-making and control across diverse domains such as strategic gameplay, robotics, and complex decision-making~\cite{sutton2018reinforcement, mnih2015human, kaufmann2023champion}. In traffic control, the ability of RL algorithms to learn from interactions with the environment makes it promising for managing the dynamic and often unpredictable nature of roadway traffic~\cite{dresner2004multiagent}. Furthermore, the development of \textit{multi-agent reinforcement learning} (MARL) provides a decentralized design perspective for distributed control systems, which avoids the scalability issues of centralized policies in systems with high dimensions or numerous agents~\cite{mannor2007multi, marl-book}.

Previous studies have applied RL and MARL to VSL control problems in simulated environments, exhibiting their potential to outperform traditional methods by adapting to evolving traffic conditions and simultaneously optimizing for multiple objectives~\cite{kuvsic2020overview}. However, applying these strategies to real-world settings presents significant challenges~\cite{han2023leveraging}. Simulations provide a controlled setting for fine-tuning algorithms, but real-world traffic introduces complexities such as diverse driver behaviors, varying vehicle types, and unpredictable weather conditions, which may influence the effectiveness of RL strategies. 
Additionally, unlike simulations that offer immediate and precise traffic state information, physical systems in the real world are subject to delays and inaccuracies, which further complicates the implementation of RL strategies~\cite{dulac2021challenges}. Therefore, deploying and evaluating RL and MARL strategies in real-world traffic scenarios is crucial to understand their practical viability and to gain insights that are unattainable in simulated environments.

The main contribution of this work is the design and assessment of the first field-deployed MARL-based VSL control system. The deployed control system consists of 67 VSL controllers (i.e., RL agents) distributed over both travel directions on a 17-mile segment of Interstate 24 (I-24) near Nashville, Tennessee, USA. Figure~\ref{fig:pipeline} provides an overview of the full deployment pipeline of the proposed MARL-based VSL control system.

Specifically, our main contributions are summarized as follows:
\begin{itemize}
    \item \textbf{We propose a design methodology for MARL-based VSL control system with large-scale real-world deployment capability:} The controllers are designed with scalability in mind under homogeneous MARL settings, accounting for commonly available traffic data, and considering multiple objectives. The optimal policy obtained during training is validated first in simulation across different operating conditions, such as traffic demand levels and driver compliance rates.
    \item \textbf{We deploy the MARL-based VSL controllers in the field:} We augment the MARL-based VSL controllers with invalid action masking and safety guards regulated by real-world constraints. The assessment results reveal that up to 98\% of the MARL-based policy's decisions directly control roadway traffic without intervention from the safety guards.
    \item \textbf{We evaluate the deployed MARL-based VSL controllers:} Compared to a previously deployed benchmark VSL algorithm, the MARL-based algorithm improves the accuracy of proactively warning about slowing traffic ahead by 14\% and reduces the response delay to non-recurrent congestion events by 75\%. Using all available preliminary data, the deployed VSL control system has reduced the crash rate and secondary crash rate by 26\% and 50\% for the controlled corridor based on a year-to-year before-after analysis. 
\end{itemize}



Compared to our preliminary  work~\cite{zhang2024field}, which primarily describes the design and deployment of the MARL-based VSL controllers, the extensions in this work are as follows. First, we present more technical details from training in simulation to the final deployment. Second, we evaluate and compare the proactive warning performance of the deployed MARL-based algorithm against the previously implemented benchmark VSL algorithm using ultra-high-resolution speed data from the I-24 MOTION traffic monitoring system~\cite{gloudemans202324}. Third, we examine and compare the response delay to non-recurrent congestion events between the deployed MARL-based algorithm and the benchmark VSL algorithm. Finally, we conduct a before-and-after analysis to evaluate the overall safety benefits of the deployed VSL control system.

To support further research, we release the MARL-based control algorithm deployed in this work along with a recorded interaction traffic dataset from a typical morning peak hour. The dataset includes state input information and the generated control outputs for each step in the algorithm. We detail each step of the execution of the deployed algorithm and demonstrate the procedure to reproduce the control actions in the codebase. The goal is to share the first field-deployed AI-based VSL control algorithm to enable further refinement and the design of novel control algorithms.

The remainder of this article is organized as follows. Section~\ref{sec:related_work} reviews VSL field deployments and recent RL-based VSL studies. Section~\ref{sec:methods} presents our pipeline from training in simulation to the ultimate deployment of the live I-24 MARL-based VSL control system. Section~\ref{sec:experiment_setup} describes the physical infrastructure details of the deployment highway segment and the software communication workflow. In Section~\ref{sec:results}, we present and discuss the behavior of the deployed MARL-based controllers as well as the traffic performance evaluation results. Finally, Section~\ref{sec:conclusions} concludes this article and provides future considerations.


\section{Related Work}
\label{sec:related_work}
Most research on variable speed limit algorithms have been tested in simulation environments to demonstrate the ability to improve traffic safety and mobility~\cite{6851128, 7006741, piao2008safety,muller2015microsimulation,abdel2006evaluation, allaby2007variable}. Field deployments of VSL systems have also been assessed, for example in the works~\cite{bertini2006dynamics, han2009best, papageorgiou2008effects, duan2012evaluating, de2018safety}. 
Due to the ease of implementation, rule-based control algorithms have been adopted by most field-deployed VSL systems~\cite{zhang2022quantifying, pu2020full, rama1999effects, long2012driver}. Although model-based control algorithms have been proposed~\cite{carlson2010optimal, yu2018optimal, han2017resolving, carlson2010optimal2}, only a few have undergone empirical testing in real-world settings. Notably, the SPECIALIST algorithm based on shock wave theory has been validated in simulation and demonstrated its capability to reduce travel time~\cite{hegyi2008specialist}. Subsequently, this algorithm was deployed on a 14 km segment of the A12 highway in the Netherlands, resolving nearly 80\% shock waves when it was activated~\cite{hegyi2010dynamic}. Another instance is the implementation of a VSL algorithm based onmodel predictive control (MPC) on Whitemud Drive in Edmonton, Canada, and the preliminary results indicated improved average travel speeds~\cite{wang2016implementation}. 

Over the past decade, RL-based approaches have gained significant attention within the traffic community~\cite{wu2021flow, zhang2024phase, vinitsky2018benchmarks, haydari2020deep}. Through experiments in a given environment to maximize cumulative reward, single-agent RL has been applied to one VSL controller~\cite{7949069, wu2020differential}, and multiple VSL controllers~\cite{zhu2014accounting, walraven2016traffic}, optimizing key traffic performance metrics such as travel time, traffic safety, and emissions. 
MARL offers advantages in multi-agent systems due to its distributed nature, which is particularly beneficial in complex engineering systems with numerous agents where centralized control could be impractical~\cite{mannor2007multi}. Zheng et al.~\cite{zheng2023coordinated} applied a MARL-based VSL control strategy to resolve consecutive bottlenecks, and the simulation results indicate significant reductions in total travel time and speed variations compared to baseline methods like independent single-agent RL and feedback control-based strategies. Fang et al.~\cite{fang2023variable} studied the effect of MARL-based VSL to reduce carbon dioxide emissions in a bottleneck located at a transition area between highways and urban roads. Other studies have investigated the application of MARL-based VSL in mixed traffic scenarios. For example, Wang et al.~\cite{wang2019new} designed a cooperative MARL-based VSL control system that enhances highway traffic mobility and safety by optimizing speed limits through \textit{vehicle-to-infrastructure} (V2I) communication. Similarly, by leveraging V2I, Han et al.~\cite{han2024multi} introduced a novel lane-specific MARL-based VSL control algorithm in a mixed traffic environment. Feng et al.~\cite{feng2024cooperative} introduced a framework integrating MARL-based \textit{connected and autonomous vehicles} (CAV) control and evolutionary-based VSL control to bridge the gap between macro and micro traffic control. In our previous work~\cite{zhang2024marvel}, we presented MARVEL (\textbf{M}ulti-\textbf{A}gent \textbf{R}einforcement-learning for large-scale \textbf{V}ariable sp\textbf{E}ed \textbf{L}imits), a MARL framework designed for large-scale VSL control with consideration of field deployment capability. A more recent review on RL and MARL-based VSL control can be found in~\cite{rhanizar2024survey}. While the above-mentioned approaches have demonstrated remarkable performance in traffic simulators, none have been deployed on real highway systems to validate their effectiveness. Simulation is an efficient way to verify a scientific hypothesis, yet real-world experiments are required to confirm the robustness and practicability of these approaches~\cite{7448922, 10571636}.

\begin{figure*}
    \centering
    \includegraphics[width=2\columnwidth]{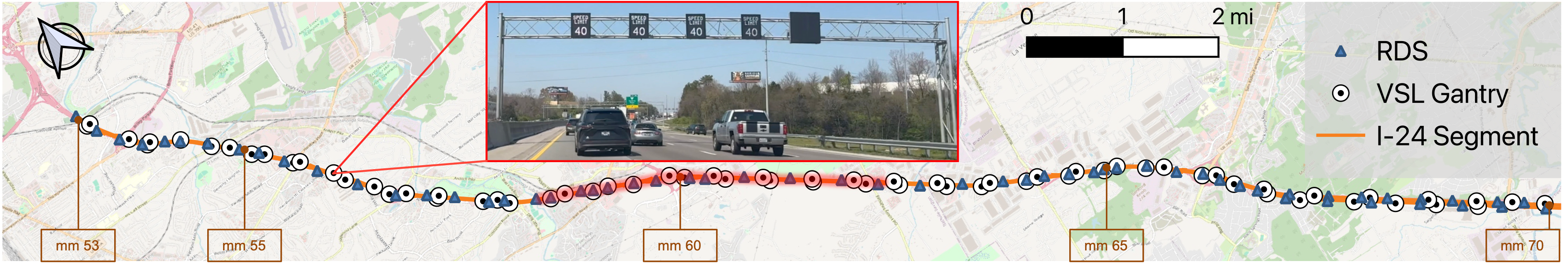}
    \caption{Map overview of the deployed MARL-based VSL control system on I-24, with the left direction heading towards downtown Nashville and the right towards Murfreesboro. The \textit{radar detection system} (RDS) units are distributed along the corridor to provide real-time traffic state data for MARL-based control algorithm. The VSL controllers are changing speed limits every 30 seconds. The overlapping segment shaded in red (from mile marker 58.7 to mile marker 62.7) is covered by I-24 MOTION, an ultra-high-resolution traffic observation system used in this study for performance evaluation. Seven VSL controllers on the westbound direction are covered by I-24 MOTION, which is used for further validation.}
    \label{fig:smartcorridor}
\end{figure*}

\section{Methods} 
\label{sec:methods}
For completeness, we briefly review the formulation of the large-scale VSL control into the MARL problem we introduced in~\cite{zhang2024marvel}, and summarize how we trained and tested a MARL policy by using environments coded in the microscopic traffic simulation software TransModeler~\cite{balakrishna2009large}, with a special focus on deployment feasibility. Additionally, we explain the real-world constraints required during deployment and our proposed solutions, which contain invalid-action masking mechanism and several safety guards. 

\subsection{Problem Formulation}
\label{sec:problem_formulation}

We consider a large-scale VSL control system where multiple VSL controllers span a long highway segment.  We formalize the MARL problem as a Stochastic (Markov) Game~\cite{marl-book}, defined as a tuple $\langle \{S^i\}_{i\in \{1,\dots, n\}}, \{A^i\}_{i\in \{1,\dots, n\}}, \{\mathcal{R}^i\}_{i\in \{1,\dots, n\}}, \mathcal{P}, n, \gamma \rangle$ for a total of $n$ agents, where $S^i$, $A^i$, $\mathcal{R}^i$ denote the observation space, action space, and reward function for agent $i$, respectively. $\mathcal{P}$ is the environment transition probability function from a given state to the next state. $\gamma$ is the discount factor used to prioritize short-term rewards over long-term rewards. The goal for each agent is to learn a policy $\pi_i(\theta_i)$ that maximizes its own cumulative discounted reward:
\begin{align}
    J^i(\theta_1, \dots, \theta_n)=\mathbb{E}_{\pi_1, \dots, \pi_n}\left[\sum_{t=0}^T \gamma^t r_t^i \right],
\end{align}
where $\theta_i$ denotes the policy parameters and $r_t^i$ is the reward of agent $i$ at time $t$. To make the final derived policies deployable, we adopt the following MARL design with the considerations of feasibility, scalability, and generalizability:

\subsubsection{Agent}
\label{sec:agent}
each VSL controller is represented by an agent. To improve the scalability of the system, we consider a homogeneous setting where all agents share the same policy parameters.

\subsubsection{Observation}
\label{sec:observation}
The observation vector is given by $[v_t^i, o_t^i, v_t^{i+1}, o_t^{i+1}, a_t^{i-1}]$, where $v_t^i$, $o_t^i$, $v_t^{i+1}$, $o_t^{i+1}$ denote the traffic speed and the traffic occupancy recorded by the traffic sensor assigned to the VSL agent (controller) $i$ and the traffic sensor assigned to the closest upstream agent $i+1$. The assigned sensor is typically located at downstream of the agent aiming to provide downstream information. $a_t^{i-1}$ represents the closest downstream agent's intended action at time $t$. We assume $a_t^{i-1}$ is the default maximum speed limit for $i=1$ (the most downstream agent). The traffic speed and occupancy that we obtain represent an average computed over a time window because of data acquisition constraints on the infrastructure side. The intuition of this design is to inform the agents of their local traffic conditions as well as the preceding (the one downstream) agent's selected action to encourage cooperation. All input features are rescaled to $[0,1]$ based on min-max normalization. 

All of these features can be acquired in real time by leveraging roadside installed traffic sensors, such as \textit{radar detection system} (RDS) used in this deployment. We note that each agent only observes the local state; it is guided by the downstream and upstream traffic characteristics in the local area without knowing traffic conditions from longer distances or further agents. The distribution of RDS units and VSL controllers at the targeted highway segment is illustrated in Figure~\ref{fig:smartcorridor}.

\subsubsection{Action}
\label{sec:action}
The policy network outputs a discrete action $a$ from the action space, defined as $A=\{30, 40, 50, 60, 70\}$, where each action represents a speed limit value (in miles per hour) that satisfies field deployment requirements.

\subsubsection{Reward}
\label{sec:reward}
The reward function accounts for drivers expectations and corridor performance. Specifically, it encompasses three terms:
\begin{align}
    r_t^i&=w_a r_{t}^{i,a} + w_s r_{t}^{i,s} + w_m r_{t}^{i,m},
\end{align}
where $r_t^{i,a}$, $r_t^{i,s}$, $r_t^{i,m}$ represent adaptability, safety, and mobility terms, respectively, and $w_a$, $w_s$, $w_m$ represent the corresponding coefficients that denote the importance of each term. The adaptability term is used to penalize an agent posting high speed limit when the traffic is a congested state. The safety term encourages the agents to coordinate so that they are able to generate a smooth slow-down speed profile upstream of the congestion tail. The mobility term encourages the agents to post a high speed limit when traffic conditions allow. For the adopted values of coefficients and the formula of each term, please refer to our previous work~\cite{zhang2024marvel}.

\subsubsection{Spatially-aware sequential decision making}
\label{sec:spatially_sequential_decision_making}
In our problem formulation, all VSL agents select their actions in a sequence, starting from the most downstream one, at any fixed time step. This mechanism is adopted to inform every agent of the preceding (downstream) agent's action, with a purpose to encourage the coordination to improve safety as well as satisfying one of the real-world constrains, which will be discussed in Section~\ref{sec:vsl_algorithm_engineering}.

\subsection{Training and Testing}
\label{sec:training_and_testing}
We conducted training and testing in the microscopic traffic simulation software TransModeler~\cite{balakrishna2009large}. TransModeler provides a Python API to set customized speed limit for different highway segments at any given time. It also allows driver compliance with the regulatory VSL system to be modeled.

\subsubsection{Training in Simulation}
The training scenario is a seven-mile long highway stretch with four lanes on I-24 westbound in Nashville, USA. To induce traffic congestion, we set up a single two-lane on-ramp merging area with a flow around 1000 veh/lane/hr. The simulation spans two hours, during which the mainstream inflow is set at 1850 veh/lane/hr for the first hour to generate congestion queue and then reduces to half to alleviate the congestion for the second hour. We implement eight VSL agents at half-mile intervals upstream of the merging area aimed at learning a cooperative policy with varying traffic conditions. This number of agents is selected based on a trial-and-error approach to balance the training efficacy and computational complexity. We set the compliance rate of 5\% for the training scenario as we expect the compliance rate on the targeted highway to be relatively low. It is important to note that the compliant vehicles would limit the ability of non-compliant vehicles to exceed the posted speed; therefore, the actual compliance rate could be much higher~\cite{zhang2022quantifying}.

We train our policy using the Multi-Agent Proximal Policy Optimization (MAPPO) algorithm~\cite{yu2022the} because of its stability during training and the relatively low number of hyperparameters to tune. For more training settings and implementation details, please refer to~\cite{zhang2024marvel}.

\subsubsection{Testing in Simulation}
We directly tested the learned policy on a 17-mile segment of I-24 modeled in TransModeler, which replicates the targeted deployment highway segment. We focus on the westbound traffic encompassing 34 VSL controllers with one traffic sensor placed 0 to 0.2 miles downstream of each VSL controller, replicating the real conditions on I-24.  We consider three testing scenarios including multiple bottlenecks and various compliance rates. Our previous results in~\cite{zhang2024marvel} demonstrate that the learned policy is able to scale to a large number of VSL agents and generalize to new environments with different traffic settings compared to the training scenario. The traffic scenarios under control of the learned policy exhibit superior mobility performance compared to a state-of-the-practice control algorithm that was initially deployed on I-24, while maintaining a lower speed variation to improve safety. 

\subsection{VSL Algorithm Engineering}
\label{sec:vsl_algorithm_engineering}
In this section, we detail the real-world constraints relevant to the intended deployment of the MARL-based VSL control algorithm. As pointed out in~\cite{dulac2021challenges}, ``reasoning about system constraints that should never or rarely be violated'' is one of the main challenges in real-world reinforcement learning applications. Here, we propose some solutions to ensure that the final algorithm adheres to the pertinent constraints.

\subsubsection{Maximum Step-Down Constraint}
The Manual on Uniform Traffic Control Devices (MUTCD) specifies a maximum permissible speed limit differential of 10 miles per hour (mph) between each pair of VSL controllers that are part of a group indicating slowdown traffic patterns~\cite{MUTCD2009}. For example, pointing at the downstream of traffic, a sequence of speed limits set at $[70, 60, 50]$ mph complies with this regulation, whereas  $[70, 50, 30]$ mph does not, as it features a differential of 20 mph, exceeding the permissible maximum. Although our safety reward term is designed to promote compliance with this constraint, violations may still occur. 

To ensure adherence to this constraint, we implement a technique known as invalid action masking (IAM)~\cite{huang2020closer}. This technique introduces a masking layer following the output of the policy network during the testing and deployment stages, which effectively avoids actions violating this constraint. With the preceding (the closest downstream) agent's intended action as part of the observation input (as described in Section~\ref{sec:observation} and~\ref{sec:spatially_sequential_decision_making}), the invalid action masking mechanism limits the action sampling to the subset of valid actions. This ensures compliance with the specified speed limit differential. The set of invalid actions for agent $i$ at time $t$ can be defined as the following:
\begin{align}
    {A_t^i}_{\text{invalid}} = \{{a}_t^i | a_t^i > a_t^{i-1} + a_{\text{diff}}, a_t^i \in A^i\},
\end{align}

where $a_t^{i-1}$ represents the intended action of the closest downstream agent $i-1$ at time $t$, and $a_{\text{diff}}$ is the maximum permissible speed limit differential for slowing down, set at $10$ mph. Note that there are no differential constraints on speed limits indicating speed-up patterns.

\subsubsection{Speed-Matching Constraint}
As an operating requirement from the \textit{Traffic Management Center} (TMC), the posted speed limits should not significantly deviate from actual traffic speeds. This requirement allows the speed limits to be easily explained to motorists, even if it prevents more exotic wave dissipation designs from being implemented.

\textit{Proposed Safety Guard:} In alignment with this requirement, we have implemented a mapping function to modify certain outputs from the learned policy. This function is designed to adjust the posted speeds to more closely reflect actual traffic conditions, particularly at the speed limit extremes. The function is defined as follows:
\begin{align}
    V=
    \begin{cases}
        \text{clip}(30, \text{min}(a_t^{i-1}+a_{\text{diff}}, f(\nu_t^i)), 70)
        &
        \text{if $a_t^i = 30$} \\
        \text{clip}(30, f(\nu_t^i), 70)
        &
        \hspace{-50pt}\text{if $a_t^i=70$ and $o_t^i\geq o_{\text{thred}}$} \\
        a_t^i
        &
        \text{Otherwise}
    \end{cases},
    \label{mappingfunction}
\end{align}
where $\text{clip}(a,\cdot,b)$ represents a clipping function that limits its output within the bounds $a$ and $b$, $f(\cdot)$ is a mapping function to map the input to the nearest multiple of 10 that is greater than the input, $o_{\text{thred}}$ is the occupancy threshold, determining when to apply this mapping when agents are selecting 70 mph. 

This mapping primarily targets adjustments at the speed limits of 30 and 70 mph, where significant deviations from actual traffic speed are most likely to occur, thus enhancing compliance with speed-matching requirements.

\subsubsection{Maximum Speed-Limit Constraint}
The maximum allowable speed limit on a highway segment is influenced by various factors, including the segment's geometric design and safety considerations. As a result, the maximum speed limits may differ across segments. For instance, while the majority of segments within the targeted field network have a maximum speed limit of 70 mph, others are capped at 65 or 55 mph. In addition, the traffic operators at the TMC may set customized maximum speed limits during special events, such as during roadway construction.

\textit{Proposed Safety Guard:} To ensure adherence to this constraint while maintaining homogeneous MARL settings for scalability, we employ a clipping function to ensure that posted speed limits do not exceed the allowable maximum for any segment. Specifically, for any generated speed limit $V$, we apply the following equation:
\begin{align}
    V'=\text{min}(V,V_{\text{max}}),
    \label{formula:maximum_constraint}
\end{align}
where $V_{\text{max}}$ represents the allowable maximum speed limit for a given segment, and $V'$ is the adjusted speed limit that satisfies this constraint.

\subsubsection{Debounce Constraint}
A \textit{bounce} is defined as a spatial sequence of speed limits at the same timestamp where all intermediate speed limits are higher than both the first and last speed limits in the sequence, which are referred to as boundary speed limits. The \textit{order} of a bounce is defined as the number of intermediate speed limits within the bounce sequence. For instance, in the direction of traffic flow, a sequence of $[30, 60, 50]$ constitutes a bounce of order $1$, whereas a sequence of $[30, 60, 50, 40]$ is a bounce of order $2$. According to local design requirements, the deployed algorithm candidates should not generate any bounce with order $1$.

\textit{Proposed Safety Guard:} To comply with this constraint, we iterate all intended speed limits to detect any bounce with an order of $1$. We apply the following equation to modify the intermediate speed limit in each identified bounce:
\begin{align}
    V''=\text{max}(V_d', V_u'),
    \label{formula:debounce}
\end{align}
where $V_d', V_u'$ represent the two boundary speed limits and $V''$ is the adjusted speed limit ensuring adherence to the debounce constraint.

\begin{figure}
    \centering
    \includegraphics[width=1\columnwidth]{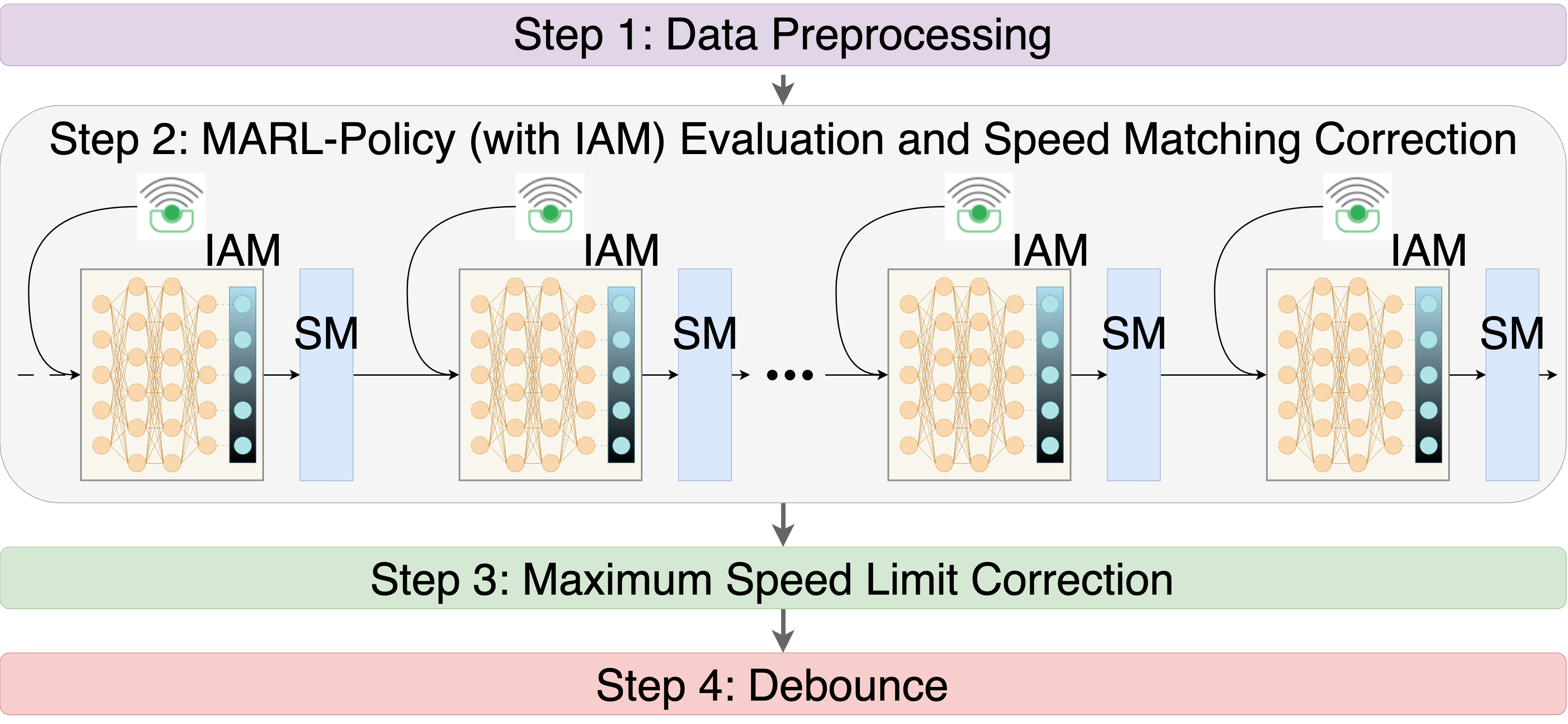}
    \caption{The deployed VSL control algorithm, centered around a MARL policy, considers all real-world constraints. IAM represents ``Invalid Action Masking'' and SM represents ``Speed-Matching''.}
    \label{fig:algorithm workflow}
\end{figure}

\subsubsection{Algorithm Integration}
This section describes the general pipeline of the deployed algorithm, from data preprocessing to the generation of final outputs. The architecture of the deployed algorithm for a set of controllers in one direction of travel is shown in Figure~\ref{fig:algorithm workflow}. This algorithm has four steps as follows:
\begin{itemize}
    \item \textit{Step 1: Data Preprocessing} -- Process all sensor data to interpolate missing values and identify the critical downstream sensor for each VSL controller. This critical sensor provides state inputs for the next step.
    \item \textit{Step 2: MARL Policy Evaluation and Speed Matching Correction} -- For each VSL controller, evaluate the MARL policy with all state inputs as described in Section~\ref{sec:methods}. With invalid action masking, the output of the policy network ensures the maximum step-down constraint. This output will go through the speed-matching module for any necessary adjustments. The updated output will then be used as a part of the state inputs to feed the upstream VSL controllers. The VSL controllers are processed in order starting from the most downstream controller, and the output of this step is a set of initial speed limits that are corrected in subsequent steps. This step is responsible for satisfying the maximum step-down and speed-matching constraints.
    \item \textit{Step 3: Maximum Speed Limit Correction} -- Process all VSL controllers (starting from the most downstream one) to make maximum speed limit corrections according to~\eqref{formula:maximum_constraint}. This step is responsible for satisfying maximum speed limit constraint.
    \item \textit{Step 4: Bounce Correction} -- Process all VSL controllers again (starting from the most downstream one) to identify if debounce constraints are violated, and correct them with the debounce logic in~\eqref{formula:debounce} to generate the final speed limits to be posted. This step is responsible for satisfying the debounce constraint.
\end{itemize}

\section{Deployment Background} 
\label{sec:experiment_setup}
This section provides a detailed overview of the deployment project, which is known as the I-24 SMART Corridor. We describe the software infrastructure supporting the implementation of the MARL-based VSL control algorithm, namely, \textit{Artificial Intelligence Decision Support System} (AI-DSS). The AI-DSS also support other decision making functionalities for I-24 SMART Corridor that are not described in this work. Lastly, we provide a brief introduction of I-24 MOTION~\cite{gloudemans202324}, an ultra-high resolution traffic monitoring system used for proactive warning evaluation in this study. 

\begin{figure*}
    \centering
    \includegraphics[width=1.3\columnwidth]{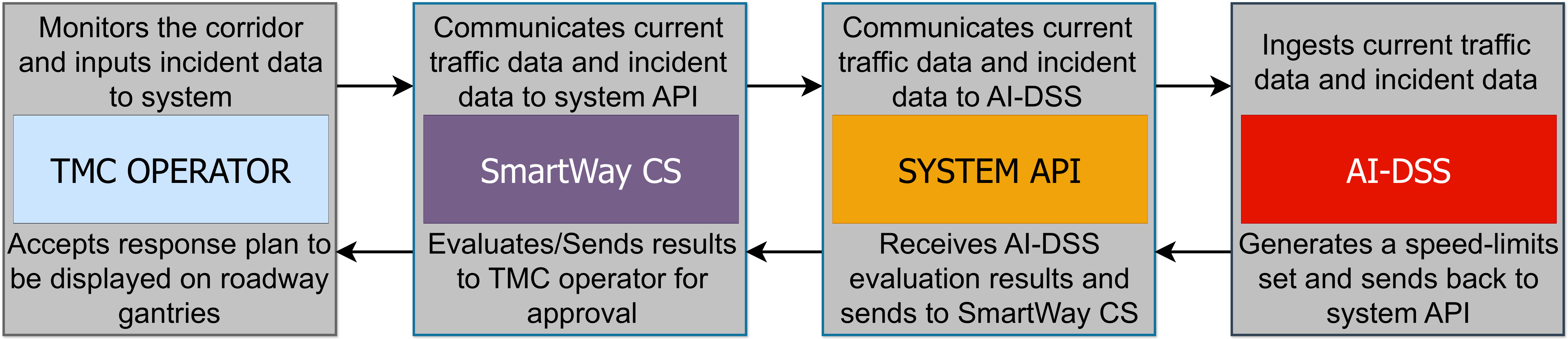}
    \caption{Overview of the communication between AI-DSS, the TMC software SmartWay CS, and the TMC operator.}
    \label{fig:ai-dss work flow}
\end{figure*}

\subsection{I-24 SMART Corridor}
\label{sec:smart_corridor}
\subsubsection{Infrastructure}
The I-24 SMART Corridor is an \textit{Integrated Corridor Management} (ICM) system run by the \textit{Tennessee Department of Transportation} (TDOT) with the goal of actively managing the increasing volumes of traffic observed on Interstate 24 south of Nashville, Tennessee. Key components of the SMART Corridor include: i.) a variable speed limit system, ii.) a lane control system (for closing lanes in response to blockages), and iii.) adjusted signal timing plans on arterial roadways to handle flow diversions when major crashes occur. The highway itself is an 8-10 lane interstate highway with an \textit{annual average daily traffic} (AADT) in excess of 160,000 vehicles. Using traffic insights from the I-24 MOTION~\cite{gloudemans202324} testbed, it is known that the highway experiences significant stop-and-go traffic daily during morning rush hour traffic.

The variable speed limit system within the I-24 SMART Corridor spans 17 miles (mile marker 53 to mile marker 70) as displayed in Figure~\ref{fig:smartcorridor}. There are 34 gantries (structures over the highway able to display the variable speed limit) on the westbound segment of the highway, and 33 gantries on the eastbound section, resulting in a spacing of approximately 0.5 miles between the gantries. The current VSL system has an operating requirement (though not a technical limitation) to post the same speed across all lanes of travel. 

\subsubsection{Radar Sensors}The sensor network that powers the VSL system is a millimeter \textit{radar detection system} (RDS), with recently upgraded Wavetronix HD sensors that have been manually calibrated for accuracy. There are 60 RDS units within the extent of the I-24 SMART Corridor (roughly 0.3 mile spacing). With a few exceptions, the devices are shoulder-mounted and measure traffic conditions across all westbound and eastbound lanes of travel. The devices are capable of measuring individual vehicle speeds per the manufacturer's documentation but are currently configured to report 30-second average speeds in each lane (for standardization with other devices statewide, where many legacy sensors exist that lack individual vehicle speed reporting capabilities). Note that the traffic measurements provided to the VSL controllers are volume-weighted pre-processed and averaged over a 90-second window to mitigate potential sensor noise. This data feed operates independently of the specific algorithmic design used for the VSL controller.  

\subsubsection{AI-DSS}
We have developed the AI-DSS, a Python-based, multi-processing software stack designed to support automated decision-making for smart highway infrastructure within regional \textit{Traffic Management Center} (TMC). Detailed descriptions of the system's architecture can be found in~\cite{van2022system}. The AI-DSS has been used to integrate the MARL-based VSL control algorithm into SmartWay CS, the production-grade active traffic management software used in TMC, enabling infrastructure deployment.

The AI-DSS is modular, allowing for intra-module upgrades and system scalability. Each module is responsible for a key function, such as handling real-time traffic data feeds communication, making and evaluating VSL control decisions, diagnosing and logging system events, and more. The architecture allows AI-DSS to be flexible with incremental growth of ICM requirements from I-24 SMART Corridor. Additionally, the multi-processing design reduces the VSL response latencies compared to a linear design, as the algorithm calculation is working simultaneously with other processes.

The communication workflow between AI-DSS and TMC is presented in Figure~\ref{fig:ai-dss work flow}. The TMC operator monitors the corridor conditions and records relevant incident information in SmartWay CS. An API in SmartWay CS allows bidirectional communications with the AI-DSS over the TCP/IP protocol using websockets. Based on the real-time traffic information from SmartWay CS, the AI-DSS implements the MARL-based control algorithm and provides the speed limits to be posted back to SmartWay CS. SmartWay CS verifies that the speed limits do not violate any constraints, and posts the speed limits to the gantries on the roadway. The AI-DSS is implemented in Python for its extensive support for libraries enabling multi-processing, websocket connectivity, database logging, and the execution of neural-network-based policies. 

\begin{figure}
    \centering
    \includegraphics[width=1\columnwidth]{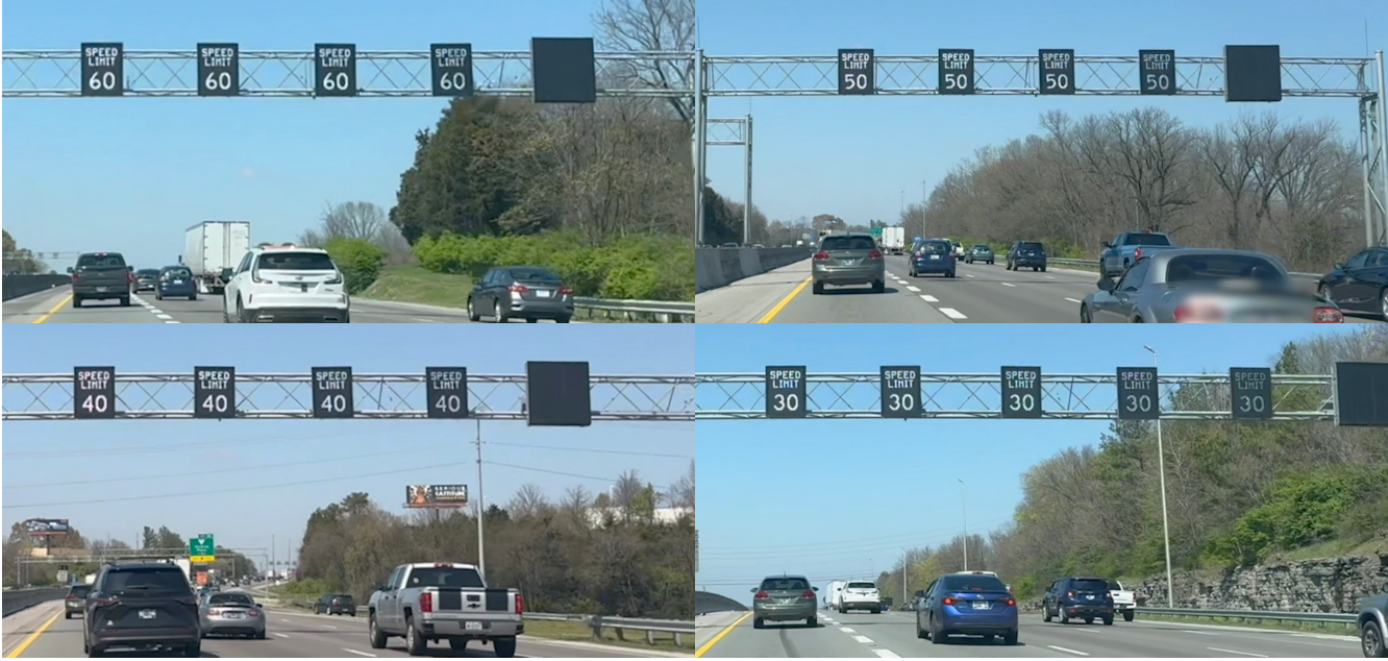}
    \caption{The deployed MARL-based VSL control system on I-24 westbound: From a driver's perspective, this figure shows four consecutive gantries that the driver encounters when approaching a congestion tail. As drivers move forward, they encounter sequentially reduced speed limits of 60 (top left), 50 (top right), 40 (bottom left), and 30 (bottom right) mph on each gantry, alerting them to the upcoming slow-down traffic patterns.}
    
    \label{fig:vsl gantries}
\end{figure}
\subsubsection{Field Testing and Deployment} \label{field testing}
Every AI-DSS version deployed in production undergoes a rigorous testing pipeline. Currently, five separate environments are designated for the AI-DSS: development, testing, production mirror at Vanderbilt; demo, and production at TDOT. Every environment has a specific configuration file to customize the websocket settings, subsystem activation, etc. The development environment is used for debugging with real-time data during the software development phase. After software development and documentation are finalized, a pre-release testing phase is conducted at Vanderbilt, lasting approximately one month. During this phase, the latest version of the VSL algorithm operates in the testing environment, where we actively monitor the controller behavior with real-time traffic data stream as input and evaluate its performance. Following this, the \textit{User Acceptance Testing} (UAT) phase begins. In this stage, the updated AI-DSS is connected to the current SmartWay CS instance and evaluated against a checklist of system responses and behaviors to ensure compatibility.
Once the updated AI-DSS version passes UAT, it is packaged and deployed in TDOT's production environment to control traffic on I-24 in real time (Figure~\ref{fig:vsl gantries}). A production mirror environment is maintained to monitor deployment status and quickly address potential system disruptions. As of December 2024, we have stabilized and executed this pipeline for 16 software releases, encompassing new system functionalities, establishment and refinements of the MARL-based VSL algorithm, integration of additional data feeds, and more.

\subsubsection{Previously Deployed VSL Algorithms on I-24}
\label{sec:old_algo}
A rule-based VSL algorithm~\cite{zhang2022quantifying} has been deployed on I-24 before the deployed MARL-based one. The core logic is to display a legal speed limit that reflects the actual traffic speed on the roadway. Several updates have been implemented since its initial deployment in order to reduce the response delay of the algorithm regarding traffic congestion and incidents. In Section~\ref{sec:results}, we will compare the performance of MARL-based VSL with the most latest version of the rule-based benchmark algorithm using empirical data collected on I-24.

\subsection{I-24 MOTION}
The Interstate 24 Mobility Technology Interstate Observation Network (I-24 MOTION) project was developed by TDOT to understand how individual vehicle interactions combine to create large-scale traffic patterns \cite{gloudemans202324}. The testbed consists of a 4.2-mile stretch of Interstate 24 within the I-24 SMART Corridor, densely monitored by 276 ultra-high definition cameras mounted on 110-foot poles. The resulting video data is processed by object detection and tracking algorithms to generate the trajectory data for individual vehicles on I-24~\cite{gloudemans2024so}. The large-scale trajectory dataset generated by I-24 MOTION can be used to understand the impact of traffic control strategies and the corresponding human behavior, providing a new opportunity to design more effective traffic control algorithms. In this study, I-24 MOTION dataset will be used to evaluate the proactive warning effectiveness of the VSL system from drivers perspective, which will be explained in Section~\ref{sec:proactive_warning}. The covered highway segment and VSL controllers by I-24 MOTION can be observed from Figure~\ref{fig:smartcorridor}.

\section{Results} 
\label{sec:results}
In this section, we present the evaluation results of the deployed MARL-based VSL control system. First, we demonstrate how the algorithm behaves during a typical morning rush hour. Second, we conduct control effectiveness and responsiveness analysis. The control effectiveness analysis reveals that 98\% of outputs from the MARL-based policy (with IAM) satisfy the operational constraints without intervention from any safety guard. The responsiveness analysis reveals that the MARL-based VSL system responds more efficiently to non-recurrent congestion events compared to both the benchmark algorithm and transportation authority agencies. Third, we evaluate the proactive warning performance with ultra-high-resolution I-24 MOTION data. The results indicate that the system outperforms the previously deployed benchmark algorithm by providing a correct warning of slow traffic ahead 88.6\% of the time (14\% improvement), with a false warning rate of 1.6\%. Finally, we conduct before-and-after analysis to evaluate the crash rate of the overall deployment. The preliminary results suggest that the VSL system has the potential to reduce crash rates compared to periods without VSL control.




\begin{figure*}
    \centering
    \includegraphics[width=2\columnwidth]{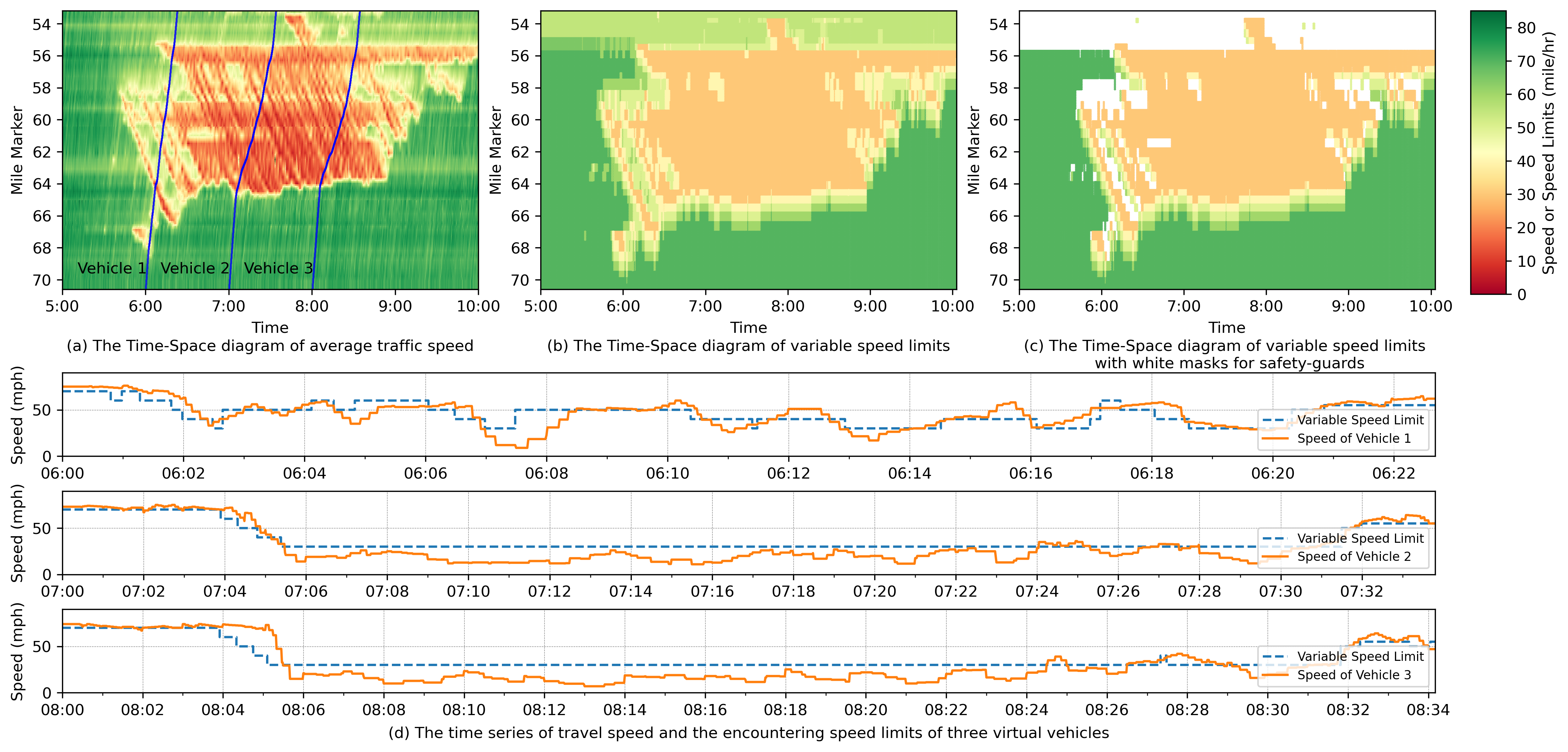}
    \caption{The MARL-based VSL control algorithm behavior from a typical morning peak hour (Monday, April 22, 2024) on I-24 westbound: (a) displays the time-space diagram of average traffic speed recorded by roadside RDS sensor in every 30 seconds. The x-axis represents time and the y-axis represents the position (in mile marker). The traffic direction is going upward along y-axis to Nashville. Three virtual vehicles are simulated starting from 6am, 7am and 8am, according to the RDS speed data, and their trajectories are overlayed on the figure. The figure (b) presents the time-space diagram of the 34 VSL gantries controlled by the MARL-based algorithm, which updates all gantries every 30 seconds. (c) shows the same diagram as (b) but with safety guards overrides masked as white. Figure (d) details the time series of the travel speed and the encountering speed limits of each virtual vehicle generated in (a).}
    \label{fig:control algorithm behavior}
\end{figure*}

\subsection{Algorithm Behavior -- examining a single peak commute period}
\label{sec:algorithm_behavior}

We begin by exploring the performance of the algorithm on a typical morning commute. Figure~\ref{fig:control algorithm behavior} (a) shows the time-space diagram of the average traffic speed of the morning peak hour of I-24 westbound on Monday, April 22, 2024. The x-axis represents time and y-axis represents mile markers of the 17-mile segment of I-24, where the traffic is going upward along y-axis to downtown Nashville. With colors denoting the traffic speed recorded by RDS sensors, Figure~\ref{fig:control algorithm behavior} (a) exhibits a typical morning rush hour congestion pattern of the selected I-24 segment, with the first congestion wave occurring at 5:30 AM near mile marker 58.

Figure~\ref{fig:control algorithm behavior} (b) displays the time-space diagram for 34 VSL gantries on I-24 westbound, which are controlled by the MARL-based VSL algorithm described in Section~\ref{sec:methods} at 30-second intervals, with the same time and space ranges as shown in Figure~\ref{fig:control algorithm behavior} (a). Note that there are six consecutive gantries closest to downtown Nashville (near MM 53-56) with a smaller maximum speed limits than the rest of the gantries. 

To understand when the MARL-based policy runs compared to the safety guards, Figure~\ref{fig:control algorithm behavior} (c) presents the same diagram as Figure~\ref{fig:control algorithm behavior} (b) but with all safety guard overrides masked in white. We note that the white part on the top of Figure~\ref{fig:control algorithm behavior} (c) is because of the six VSLs with smaller maximum speed limits (which breaks the homogeneous agents assumption in the design), for which the Maximum Speed Limit Correction safety guard has been triggered.

To understand how the deployed algorithm behaves from the perspective of a driver, we generate trajectories of three simulated vehicles traveling according to the RDS speed data as shown in Figure~\ref{fig:control algorithm behavior} (a), using the method of~\cite{ji2024virtual}. 

Figure~\ref{fig:control algorithm behavior} (d top, middle, and bottom) show the time series of the virtual vehicles based on the prevailing traffic speed (orange), and the posted speed limits (dashed blue). Figure~\ref{fig:control algorithm behavior} (d top) shows Vehicle 1 encountering multiple stop-and-go waves which are visualized in the oscillations in orange. These speed fluctuations are occasionally in excess of 50 mph, which create the opportunity for high speed rear end collisions. Figure~\ref{fig:control algorithm behavior} (d top) shows the VSL speed limits drop in advance of the vehicle encountering slower traffic, providing a proactive warning to the driver. 

Vehicles 2 (Figure~\ref{fig:control algorithm behavior} (d middle)) and 3 (Figure~\ref{fig:control algorithm behavior} (d bottom)) start later than Vehicle 1, and encounter more congestion, resulting in a longer travel time. Prior to the vehicles encountering congestion, they receive correct warnings to slow down, indicating the system is working to give correct warnings of slow traffic ahead. This can best be observed at 7:04 am--7:06 am for Vehicle 2 in Figure~\ref{fig:control algorithm behavior} (d middle), and from 8:04 am--8:06 am for Vehicle 3 in Figure~\ref{fig:control algorithm behavior} (d bottom).

\begin{table}
\caption{The daily effectiveness percentage (AVG$\pm$STD) of MARL-Policy with IAM (Policy), Speed-Matching (SM), Maximum Speed Limit Correction (MSLC), and Debounce (DB). ``I'', ``E'' categorize those gantries with custom max speed limit being included/excluded. ``WB'' and ``EB'' refer to ``Westbound'' and ``Eastbound'', and ``PH'' refers to ``Peak Hour''.} 
  \centering
  \begin{tabularx}{\columnwidth}{@{}llcccc@{}} 
    \toprule
    & Dataset     & Policy (\%)  & SM (\%) & MSLC (\%) & DB (\%) \\ 
    \midrule
    \multirow{4}{*}{\begin{tabular}[c]{@{}l@{}}I\end{tabular}} 
    & I-24 WB      &81.3$\pm$0.8    &1.8$\pm$1.1      &16.1$\pm$1.2  &0.8$\pm$0.5    \\
    & I-24 WB PH &78.4$\pm$2.7    &7.3$\pm$2.2     &10.7$\pm$2.4  &3.6$\pm$1.0  \\
    & I-24 EB      &87.4$\pm$0.9    &2.6$\pm$1.6      &9.6$\pm$1.3   &0.4$\pm$0.2 \\
    & I-24 EB PH &84.4$\pm$3.2    &12.6$\pm$3.4      &1.2$\pm$2.0   &1.8$\pm$0.8 \\ 
    \midrule
    \multirow{4}{*}{\begin{tabular}[c]{@{}l@{}}E\end{tabular}} 
    & I-24 WB      &98.4$\pm$1.1    &1.3$\pm$0.9      &0    &0.3$\pm$0.3\\
    & I-24 WB PH   &93.0$\pm$2.5    &5.2$\pm$1.9      &0    &1.8$\pm$0.7\\
    & I-24 EB      &97.6$\pm$1.7    &2.1$\pm$1.4      &0    &0.3$\pm$0.2\\
    & I-24 EB PH   &86.5$\pm$4.5    &11.7$\pm$3.7      &0    &1.8$\pm$0.9\\ 
    \bottomrule
  \end{tabularx}
\label{table:percentage}
\end{table}

\subsection{Control Effectiveness and Responsiveness -- examining with six months operating data}
Next we look at six months of operations data to determine how frequently the MARL policy runs compared to the safety guards, and to determine the system responsiveness to non-recurrent congestion.
\subsubsection{Control Effectiveness}
\label{sec:control_effectiveness}
We explore dataset collected from March 8, 2024 to September 8, 2024, encompassing 34,032,683 decisions across 67 gantries. Table~\ref{table:percentage} presents the percentage of decisions where the MARL-based policy with IAM (Policy) directly takes control, those where Speed-Matching (SM) intervenes to correct the Policy, and those involving final adjustments through Maximum Speed Limit Correction (MSLC) and the Debounce (DB) logic.

The MARL policy runs 81.3\% of the time on I-24 Westbound (WB) and 87.4\% on I-24 Eastbound (EB) daily, across all 67 VSL gantries. The maximum speed limit correction is the largest cause of violation, running 16\% (WB) and 9\% (EB) respectively. Again, this is because the agents are assumed to be homogeneous in training, but six gantries in each direction have lower maximum allowable speed limits, violating the homogeneous assumption. Consequently MSLC is active anytime traffic is light and maximum speed limits are used. When the statistics are recomputed using all gantries with a maximum speed limit of 70 mph, we find that the MARL policy runs 98\% (WB) and 97\% (EB). The speed matching correction also runs about 1-2\% of the time, which is needed to prevent the algorithm from posting something significantly different from prevailing traffic conditions.

\subsubsection{Control Responsiveness to Non-Recurrent Events}
\label{sec:vsl_response_time_to_events}


To understand the response delay of the deployed MARL-based VSL system, we manually labeled all non-recurrent congestion events from RDS-based time-space diagrams. Non-recurrent congestion can be caused by crashes or other incidents, and the fast response of the VSL system can be used to warn drivers of atypical traffic, preventing further crashes or incidents.

As mentioned in Section~\ref{sec:smart_corridor}, the RDS units, distributed approximately every 0.3 miles, provide sufficiently fine-granularity to identify non-recurrent traffic congestion. We generate time-space plots similar to ~\ref{fig:control algorithm behavior} (a) for each day and in each direction, and then label all non-recurrent congestion events in these plots. This includes the start and end times and the location of the event.  We identify 546 non-recurrent congestion events, with 337 of them having corresponding records in the \textit{Traffic Management Center} (TMC) database. Not all non-recurrent congestion generates a record in the TMC, but when the record is available, we can compare the delay of the VSL system as well as the reporting delay in the TMC.

The labeled dataset covers the period from January 1, 2024 to June 20, 2024. From January 1-March 8, the MARL policy was deployed in an open-loop test environment, allowing the analysis during that period to still be conducted. After March 8, the MARL policy ran closed-loop on the live system.

\begin{figure}
    \centering
    \includegraphics[width=1\columnwidth]{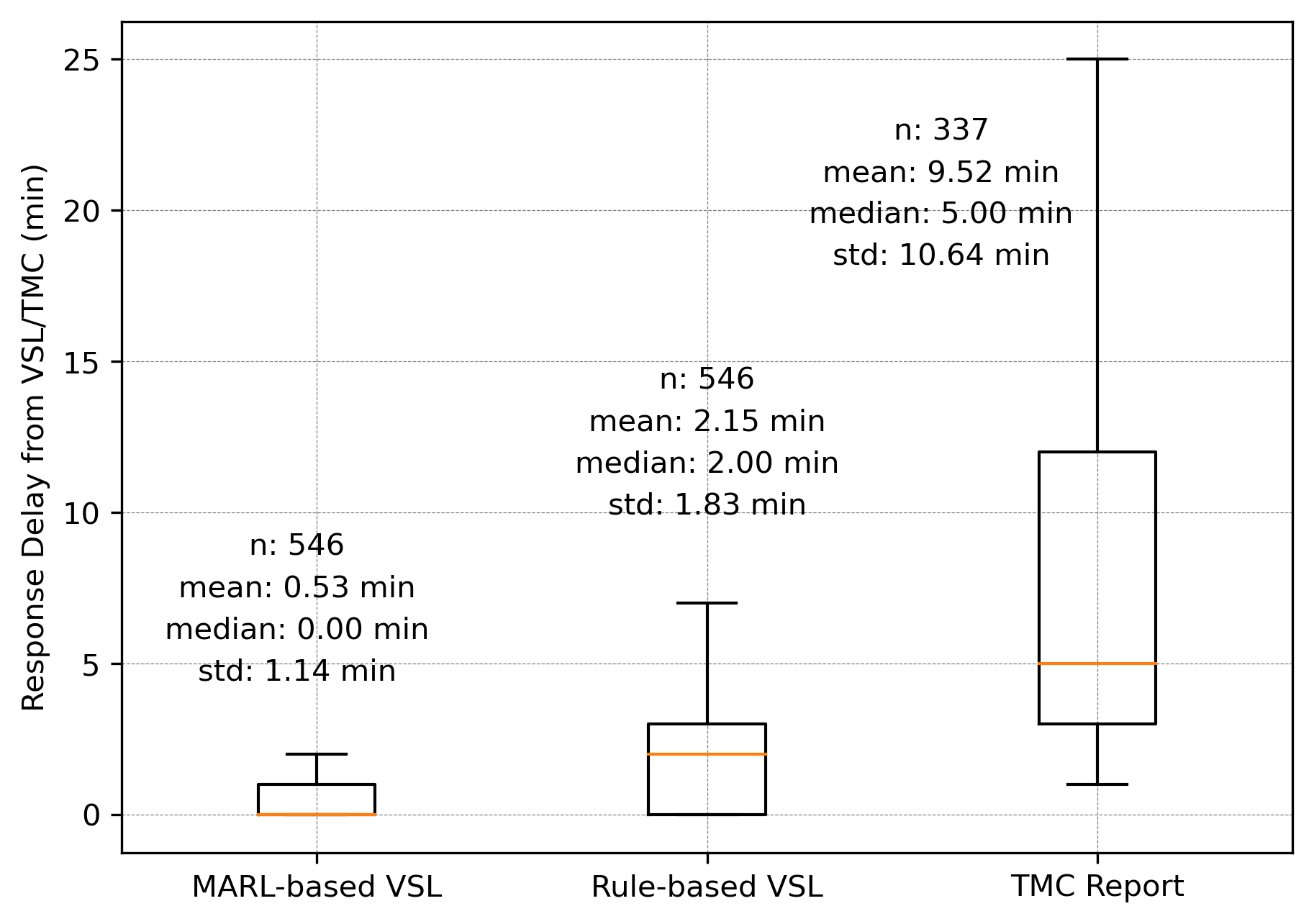}
    \caption{The boxplot of VSL response delay and TMC report delay regarding the estimated non-recurrent congestion start time. The variable $n$ represents the number of samples in each dataset. The MARL-based VSL system response time is on average 9 minutes ahead of the TMC report (75\% faster than the benchmark algorithm, and 90\% faster than the TMC).}
    \label{fig:vsl_response}
\end{figure}

Figure~\ref{fig:vsl_response} shows the boxplot of response delay of the MARL VSL controller and the reporting delay in the TMC. The MARL policy responds on average less than one minute after the event start time, with a standard deviation of 1.14 minutes. In contrast, the report in the TMC database is more than nine minutes after the event start, with a standard deviation of 10 minutes. The MARL policy is also more responsive than the previously deployed rule-based algorithm (see Section~\ref{sec:old_algo}) which typically responds in 2.15 min with a standard deviation of 1.83 min. In short, the VSL system actively responds to non-recurrent congestion events nearly instantaneously, often before the TMC learns of the event. This pattern has led some TMC staff to use the VSL activations as a type of incident detection algorithm, where nearby cameras are used to confirm incidents when VSL runs at atypical times or locations.



\subsection{Proactive Warning -- examining with I-24 MOTION}
\label{sec:proactive_warning}
In this section, we compare two algorithm settings by analyzing their responsiveness accuracy in response to downstream slow-down traffic. 
\subsubsection{Proactive VSL Preliminaries}
VSL systems can be viewed as a type of proactive warning system in which the reduction in speed limits inform drivers that downstream traffic ahead is moving slowly, giving drivers time to adjust their speed before encountering congestion~\cite{corthout2010assessment}. The proactive warning correctness of the VSL system is evaluated from the perspective of individual vehicles~\cite{10571636, 7448922}. By using a fine-grained speed field, virtual vehicle trajectories are generated to mimic the behavior of individual vehicles driving through the highway corridor. These trajectories, when combined with the VSL logs, enable the determination of speeds for virtual vehicles and corresponding speed limits at specific times and locations. 

To evaluate the proactive warning performance of the VSL system, we first introduce the following definitions, with subsequent definitions of successful warning rate and false warning rate, as established in~\cite{corthout2010assessment}:
\begin{itemize}
    \item \textbf{\textit{Situation to be Warned:}} This situation arises when, between two consecutive VSL gantries, a vehicle travels below the minimum speed limit. When this occurs, the upstream VSL gantry should display the minimum speed limit as the vehicle passes the gantry.
    \item \textbf{\textit{Successful Warning:}} For any \textit{Situation to be Warned}, the upstream VSL gantry is expected to display the minimum speed limit at the time a virtual vehicle passes. Failing to do this is considered as a \textit{Missed Warning}.
    \item \textbf{\textit{Warning:}} A VSL gantry posts the minimum speed limit.
    \item \textbf{\textit{False Warning:}} For any \textit{Warning}, the lowest speed in the downstream trajectory segment always exceeds the minimum speed limit by more than an allowable maximum deviation. 
\end{itemize}

Considering all of the \textit{situations to be warned} (i.e., traffic is slower than 30 mph), the \textit{Successful Warning Rate} (SWR) is:

\begin{align}
    \textbf{\textit{SWR}} = \frac{\text{\# of \textit{Successful Warning}}}{\text{\# of \textit{Situation to be Warned}}},
\end{align}
and the \textit{missed warning rate} is the complement of SWR.  
    
    Considering all of the times a warning is posted (i.e., a gantry posts 30 mph), the \textit{False Warning Rate} (FWR) is defined as:
\begin{align}
    \textbf{\textit{FWR}} = \frac{\text{\# of \textit{False Warning}}}{\text{\# of \textit{Warning}}}.
\end{align}

\begin{figure}
    \centering
    \includegraphics[width=1\columnwidth]{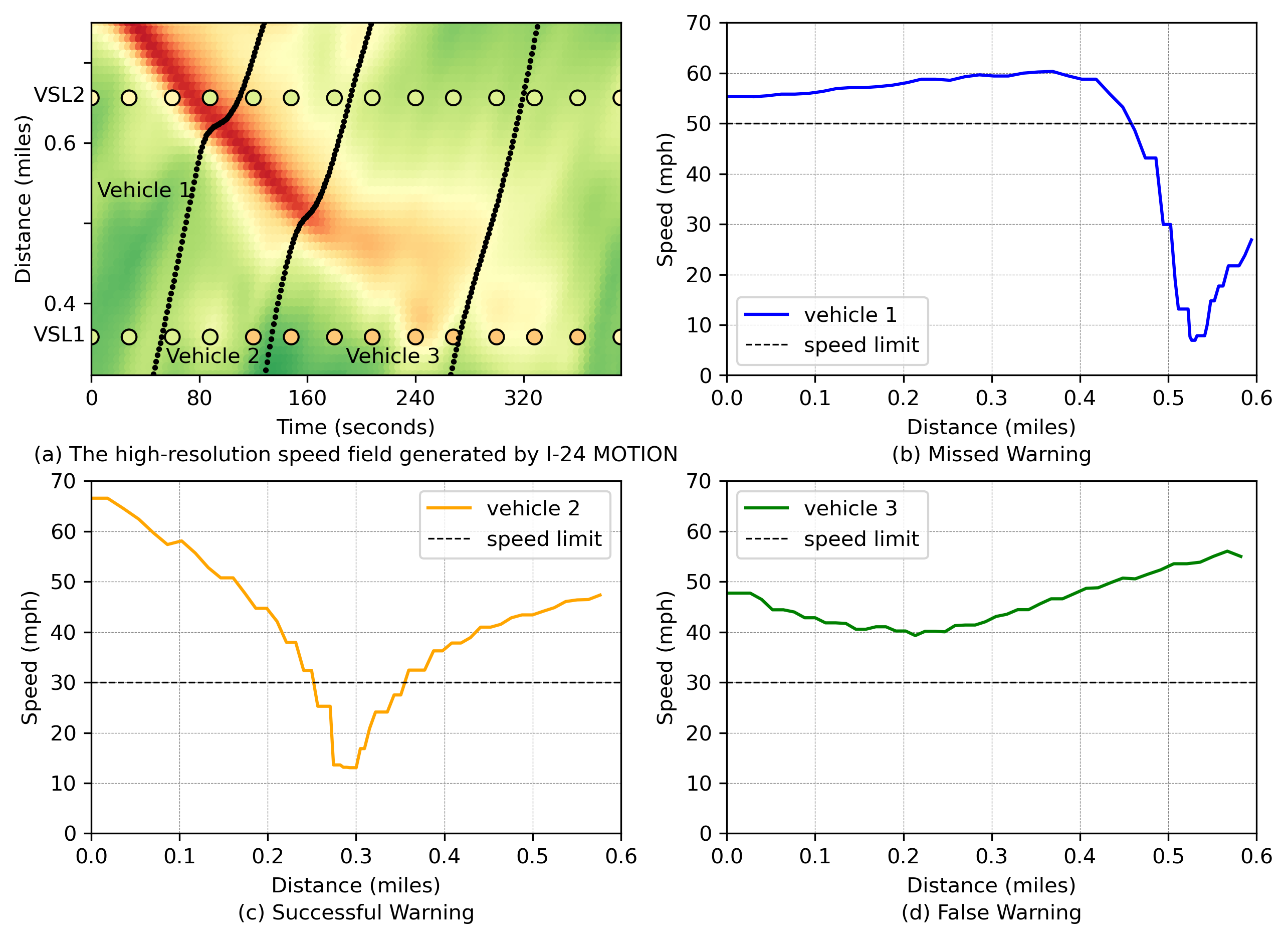}
    \caption{An illustration of our definitions of \textit{Missed Warning}, \textit{Successful Warning}, and \textit{False Warning}: (a) This time-space diagram contains a speed-field spanning two consecutive VSL gantries for 400 seconds. The color represents different speeds or speed limits generated by the VSL algorithm. Three virtual vehicle trajectories are generated for illustration. (b) The driving speed and the corresponding speed limit of vehicle 1 from VSL1 to VSL2. (c) The driving speed and the corresponding speed limit of vehicle 2 from VSL1 to VSL2. (d) The driving speed and the corresponding speed limit of vehicle 3 from VSL1 to VSL2.}
    \label{fig:illustrate}
\end{figure}

On our deployment 30 mph is the minimum speed limit used to define the \textit{Situation to be Warned} and 10 mph is used as  the maximum allowable deviation for a \textit{False Warning}. Figure~\ref{fig:illustrate} provides an illustration for a \textit{Successful Warning}, a \textit{Missed Warning}, and a \textit{False Warning}. Figure~\ref{fig:illustrate} (a) displays three virtual vehicle trajectories along with the speed limits of two consecutive gantries. Figure~\ref{fig:illustrate} (b), (c), (d) present the speed profile between two VSL gantries and the corresponding speed limit when they pass VSL1 for vehicle 1, vehicle 2, and vehicle 3, respectively. Vehicle 1 passes VSL1 with a posted speed limit of 50 mph and eventually approaches a traffic wave with a minimum speed below 10 mph before it reaches VSL2. This 50 mph posted by VSL1 is considered a Missed Warning (because Vehicle 1's speed dropped below 30 mph, a speed limit of 30 mph should have been posted).  On the contrary, vehicle 2 passes VSL1 with a posted speed limit of 30 mph as a warning message and it indeed encounters lower speed in the downstream traffic. This 30 mph posted by VSL1 is considered as a \textit{Successful Warning}. Finally, vehicle 3 gets a warning message of 30 mph when passes VSL1 but keeps a speed higher than 40 mph during the downstream travel. This 30 mph posted by VSL1 is considered as a \textit{False Warning}.

\begin{table}
\centering
\footnotesize
\caption{Experiment design: five experiments in total accounting for the impact of algorithm, control input, and evaluation data source.}
\label{table:experiment_design}
\begin{tabularx}{\columnwidth}{cccc} 
\toprule
Experiment & Algorithm & Control Input  & Evaluation Data Source \\ \hline
1    & MARL    & RDS       & RDS          \\
2    & MARL    & RDS       & MOTION      \\  
3    & Rule    & RDS       & MOTION      \\
4    & MARL    & \makecell{MOTION \\ (lane-average)}   & MOTION      \\
5    & MARL    & \makecell{MOTION \\ (lane-specific)}  & MOTION     \\
                                 \bottomrule
\end{tabularx}
\end{table}

\begin{table}
\centering
\footnotesize
\caption{Proactive Warning metrics for all experiments. Best result in bold. * indicates metrics are computed via an approximate method.}
\label{table:result}
\begin{tabularx}{\columnwidth}{ccccccc} 
\toprule
Experiment & Metric & Lane1 & Lane2 & Lane3 & Lane4 & Overall \\ \hline
\multirow{2}{*}{1}  & SWR* (\%)    &97.6    &97.9      &98.1   &98.1    &98.0    \\
                    & FWR* (\%)    &11.7    &7.3       &5.7    &7.2     &8.0     \\  
\multirow{2}{*}{2}  & SWR (\%)    & 88.3   & 88.1   & 88.5   & 89.7   & 88.6    \\
                    & FWR (\%)    & 2.7    & 1.5    & 1.1    & 1.6    & 1.6     \\
\multirow{2}{*}{3}  & SWR (\%)    & 74.4   & 73.7   & 73.7   & 75.8   & 74.4    \\
                    & FWR (\%)    & 2.2    & 1.4    & 1.0    & 1.6    & 1.7     \\
\multirow{2}{*}{4}  & SWR (\%)    & 94.3   & 93.8   & 93.4   & 94.5   & 94.0    \\
                    & FWR (\%)    & \textbf{1.3}    &\textbf{0.7}    & \textbf{0.5}    & \textbf{0.9}    & \textbf{0.8}     \\ 
\multirow{2}{*}{5}  & SWR (\%)   & \textbf{96.1 }  &\textbf{ 96.2}   & \textbf{96.2}   & \textbf{95.9}   &\textbf{96.1}                       \\
                    & FWR (\%)   & 2.2    & 1.5    & 1.3    & 1.5    & 1.6     \\
                                 \bottomrule
\end{tabularx}
\end{table}

\subsubsection{Proactive Warning Analysis Questions and Methods}
 We consider the following analysis questions:
\begin{itemize}
    \item \textbf{Q1}: How does data granularity in evaluation data sources affect the evaluation results given a specific control algorithm and a fixed type of control input?
    \item \textbf{Q2}: How does the MARL-based VSL system perform compared to a previously implemented rule-based control algorithm?
    \item \textbf{Q3}: How would different levels of data granularity in control inputs affect system performance?
\end{itemize}
To answer these questions, we design five experiments as presented in Table~\ref{table:experiment_design}, with two VSL control algorithms running in parallel. The first is the currently deployed MARL-based algorithm, and the second is a rule-based benchmark algorithm previously deployed on I-24 (\ref{sec:old_algo}).

We use the log files of the VSL control system and the fine-grained speed field from I-24 MOTION system for the analysis. The data covers the morning peak hours (6 AM to 10 AM) of I-24 westbound within I-24 MOTION segment from 15 weekdays in June 2024. The VSL logs contain all the operational information, including the measured RDS speeds and posted speeds, which are recorded at 30-second intervals. There are a total of 50,400 VSL logs for each algorithm during the above-mentioned periods. The raw trajectories from I-24 MOTION dataset of each lane were first converted to a lane-level smoothed macroscopic speed field~\cite{gloudemans202324,ji2024virtual,ji2024stop} with a spatio-temporal resolution of 4 seconds and 0.02 miles across the entire four-mile stretch of I-24. Virtual trajectories were then generated on the smoothed macroscopic speed field in a 15-second interval for each lane following the method in~\cite{ji2024virtual}, allowing analysis from the perspective of an individual vehicle.

For \textbf{Q1}, \textit{experiments 1 and 2} use the MARL-based VSL control algorithm with RDS as the control input but employ different evaluation data sources to analyze the impact of evaluation data granularity on the results. For \textbf{Q2}, we apply I-24 MOTION dataset to evaluate both control strategies with RDS as control input (\textit{experiments 2 and 3}). We note that the rule-based algorithm was not implemented in the field, but ran in open loop mode without affecting traffic. For \textbf{Q3}, we replace low-resolution RDS data with ultra-high-resolution I-24 MOTION data as the input to the VSL controller. In \textit{experiment 4}, we replace the RDS data averaged across all lanes as the speed input with I-24 MOTION speed averaged across all lanes. In \textit{experiment 5}, we consider a lane-level VSL control algorithm where speed limits could differ across lanes; specifically, we replaced the RDS speed averaged across all lanes with lane-specific speed data recorded by I-24 MOTION.

\subsubsection{Results}
The results of each experiment are displayed in Table~\ref{table:result}. 

 First, we note that the evaluation results vary significantly depending on the evaluation data granularity. From experiment 1 (where the coarser grained RDS data is used as both the control input and evaluation datasource), one might mistakenly conclude that the MARL-based VSL control algorithm achieves excellent proactive warning performance. However, this result appears overly optimistic when compared to experiment 2 (which uses RDS data as control input but higher resolution I-24 MOTION data for the evaluation). For example, the successful warning rate drops by nearly 10\% when using the higher resolution data for evaluation.
    
Using I-24 MOTION data for the evaluation, experiments 2 and 3 show that the MARL-based VSL algorithm outperforms the traditional rule-based algorithm in terms of SWR (88.6\% vs 74.4\%) and FAR (1.6\% vs 1.7\%). This highlights the ability of the MARL-based approach to adapt to the dynamic variability of traffic on this roadway.

Experiment 4 demonstrates that the MARL performance could be improved (e.g., SWR of 94.0\% up from 88.6\%) simply by switching the control input data from RDS data to I-24 MOTION, or to another data source with less spatio-temporal averaging. Alternatively, performance could be improved with a traffic estimation layer applied to the RDS data (to approximate the resolution of I-24 MOTION) which would be more scalable in practice.

Finally, experiment 5 demonstrates a modest improvement in the SWR can be hypothetically achieved (c.f., experiment 4) by using lane-specific VSL messages if the restriction to post the same speed limit across all lanes is relaxed. This is offset by a slight increase in the FWR.

\begin{table}
\centering
\footnotesize
\caption{The crash rate before-after analysis depending on whether VSL was active when the crash occured.}
\label{table:crash_rate}
\begin{tabularx}{\columnwidth}{cccc} 
\toprule
Period & \makecell[c]{Crash Rate \\ with Active VSL \\ (crashes/month)} & \makecell[c]{Crash Rate \\with Inactive VSL \\ (crashes/month)}  & \makecell[c]{Secondary Crash Rate \\ (crashes/month)} \\ \hline
Before    & 18.4    & 23.6       & 7.2          \\
After     & 15.8 ($\downarrow$ 14\%)    & 26.5 ($\uparrow$  12\%)       & 3.6 ($\downarrow$ 50\%)     \\  
                                 \bottomrule
\end{tabularx}
\end{table}

\subsection{Before-and-After Analysis -- examining crash rate of two years}
\label{sec:before_after_analysis}
Here we present a preliminary crash analysis results, to gain insights into the safety performance since the launch of the VSL system in June 2023. Evaluation over longer time horizons is part of an ongoing evaluation.

To evaluate the crash reduction performance, it is important to note that the VSL system runs 24 hours a day. During free-flow conditions, it (should) post the maximum speed limit, and when it does, we consider the system in an \textit{inactive} state. We consider the VSL system \textit{active} any time that the posted speed limit is lower than the maximum speed limit. This distinction is important, since the VSL system cannot prevent crashes (e.g., through proactive warning) when it is in the inactive state. As shown in the previous section, the system provides high quality proactive warnings with few false warnings.


To evaluate the effectiveness of the VSL system, we focus on crashes that occur when the system is active. For the year following the VSL implementation on June 20, 2023, we use the VSL logs at the time and location of each crash to determine whether the nearest VSL controller is active. Note that during this time, the MARL and rule based algorithm are both active over a portion of the time. We bundle the data together to boost the overall sample size and to avoid seasonality effects.  We note that the Annual Average Daily Traffic (AADT) has increased by 6.6\% when comparing the year prior to VSL deployment to the year after.

For the year prior to the VSL implementation, we reprocess the RDS data through the VSL algorithm to hypothetically determine if the VSL would have been active, had it been deployed. Similar to the methodology described previously in Section~\ref{sec:vsl_response_time_to_events}, we manually labeled all non-recurrent congestion from RDS-based time-space diagrams for two years---one year before and one year after VSL went live. We collect those non-recurrent congestion events that correspond with crashes recorded in the TMC database.


Table~\ref{table:crash_rate} presents the preliminary before-and-after crash analysis results. We observe that the crash rate decreases from 18.4 crashes per month to 15.8 crashes per month—a 14\% reduction—when the VSL was (or would have been, in the before data) active. Conversely, when the VSL was inactive, the crash rate increases from 23.6 crashes per month to 26.5 crashes per month, reflecting a 12\% increase. These findings suggest that without the deployment of VSL, the total crash rate on I-24 would likely have gone up (similar to other corridors nearby).

We also conduct a secondary crash analysis since the VSL system has the potential to prevent these incidents by providing slower speeds to warn upstream traffic. The results show that the secondary crash rate decreases from 7.2 crashes per month to 3.6 crashes per month—a 50\% reduction. We reemphasize that the preliminary crash statistics are likely to change as more data continues to be collected, but the preliminary results are promising. 


\section{Conclusion}
\label{sec:conclusions}
This work presents the first MARL-based VSL system deployed in the real world on the I-24 highway near Nashville, Tennessee, which continues to operate today. We describe the deployment pipeline, including training and testing in simulation, the algorithm engineering to involve real-world constraints, the open-loop testing with real-time traffic data streams, and the final closed-loop field deployment.

We conduct controller-level analysis and demonstrate the system behaviors during highway peak hours. Our effectiveness analysis reveals the feasibility to deploy a simulation-based MARL policy in real-world settings with safety guards. We find that the safety guards only run a small portion of time compared to the MARL policy. In comparison to the previously deployed benchmark VSL algorithm on I-24, the MARL-based algorithm provides more accurate warning information to drivers about the downstream slow-down traffic patterns. Furthermore, MARL-based VSL system achieves an efficient response performance in response to non-recurrent congestion events, outperforming both the benchmark algorithm and highway emergency response teams. Overall, we observe a decreased crash rate during the time period after VSL implementation when the VSL system was or would have been posting slow speeds. 

 In future work, we are interested in conducting further experiments to promote higher compliance rates among drivers. This could be pursued by applying V2I techniques to certain amount of automated vehicles to follow the speed limits, promoting the local compliance from surrounding vehicles. Additionally, distilling the MARL-based policy can be useful to understand the internal control logic and provide insights into the general VSL algorithm design. 


\section*{Acknowledgments}
The authors express their gratitude to the Tennessee Department of Transportation (TDOT), Southwest Research Institute (SwRI), Arcadis, and Stantec for their assistance in the pioneering AI-based VSL deployment on Interstate 24. We also thank Caliper for their technical support with TransModeler, and Caleb Van Geffen and Josh Scherer for their contributions to the development of AI-DSS during their time as students at Vanderbilt University. The content of this report reflects the views of the authors, who are responsible for the facts and accuracy of the information presented. The U.S. Government assumes no liability for the contents or use thereof. This research was supported by a U.S. Department of Transportation Grant (No. 693JJ22140000Z44ATNREG3202), a National Science Foundation Grant (No. CNS-2135579), and a Tennessee Department of Transportation Grant (No. RES2023-20).

\bibliographystyle{ieeetr}
\bibliography{main}

\vfill

\end{document}